\documentclass[aps,prl,twocolumn,showpacs,preprintnumbers,amsmath,amssymb]{revtex4}
\usepackage{graphicx}
\usepackage{graphicx}
\usepackage{dcolumn}
\usepackage{bm}

\newcommand{\be}{\begin{eqnarray}}
\newcommand{\ee}{\end{eqnarray}}

 \newcommand{\gsim}{\mathrel{\hbox{\rlap{\lower.55ex \hbox {$\sim$}}
                   \kern-.3em \raise.4ex \hbox{$>$}}}}
\newcommand{\lsim}{\mathrel{\hbox{\rlap{\lower.55ex \hbox {$\sim$}}
                   \kern-.3em \raise.4ex \hbox{$<$}}}}

\def\roughly#1{\mathrel{\raise.3ex\hbox{$#1$\kern-.75em%
\lower1ex\hbox{$\sim$}}}}
\def\lsim{\roughly<}
\def\gsim{\roughly>}

\begin{document}


\title{Angular Dependence of Jet Quenching Indicates\\
Its Strong Enhancement Near the QCD Phase Transition}

\author{Jinfeng Liao$^{1,2}$}\email{jliao@lbl.gov}
\author{Edward Shuryak$^1$}\email{shuryak@tonic.physics.sunysb.edu}
\affiliation{$^1$Department of Physics and Astronomy, State
University of New York, Stony Brook, NY 11794\\
$^2$Nuclear Science Division, Lawrence Berkeley National
Laboratory, Berkeley, CA 94720}

\date{\today}

\begin{abstract}
We study dependence of jet quenching on matter density, using
``tomography'' of the fireball provided by RHIC data on azimuthal
anisotropy $v_2$ of high $p_t$ hadron yield at different
centralities. Slicing the fireball into shells with constant
(entropy) density, we derive a ``layer-wise geometrical limit''
$v_2^{max}$ which is indeed above the data $v_2<v_2^{max}$.
Interestingly, the limit is reached only if quenching is dominated
by shells with the entropy density exactly in the near-$T_c$
region. We show two models that simultaneously describe the high
$p_t$ $v_2$ and $R_{AA}$ data and conclude that such a description
can be achieved only if the jet quenching is few times stronger in
the near-$T_c$ region relative to QGP at $T>T_c$. One possible
reason for that may be recent indications that the near-$T_c$
region is a magnetic plasma of relatively light color-magnetic
monopoles.
\end{abstract}


\maketitle

{\em Introduction.---} Recent experiments at the Relativistic
Heavy Ion Collider (RHIC) are dedicated to  study possible new
forms of QCD matter, with increasing  energy density. In such
collisions the produced matter equilibrates as Quark-Gluon Plasma
(QGP)\cite{Shuryak:1978ij} and then cools down through the
near-$T_c$ (M) phase (M for
mixed/median/magnetic\cite{Liao:2006ry}) into the usual hadronic
phase (H). To probe the created matter in an externally
controllable way, like using X-ray for medical diagnosis
 is impossible. However, high energy jets are internal probes:
 propagating through the fireball, they
 interact  -- and thus obtain important information
about the medium  -- as proposed long ago in
Refs\cite{Bjorken:1982tu,Appel_Blaizot_McLerran,Gyulassy:1990ye}.
 In heavy ion collisions this energy loss can be
manifested in the suppression of observed hadron spectra at high
transverse momenta $p_t$, as well as in the suppression of
back-to-back di-hadron correlations with a high-$p_t$ trigger, when
compared with pp and d-A collisions. The ``jet quenching''
phenomenon  is one of the major discoveries by the RHIC
experimental program\cite{rhic_white_paper}.

The suppression  is quantified by comparison of the
inclusive spectra $d^2 N^{AA}/dp_td\eta$ in ion-ion(AA) collision
to a nucleon-nucleon(pp) reference $d^2 \sigma^{NN}/dp_td\eta $
via the Nuclear Modification Factor $R_{AA}(p_t)$. :
\begin{equation}\label{eqn_raa}
R_{AA}(p_t) \equiv \frac{d^2 N^{AA}/dp_td\eta}{T_{AA}\cdot d^2
\sigma^{NN}/dp_td\eta }
\end{equation}
with $T_{AA}$ the nuclear overlap function which scales up single
NN cross section to AA according to expected number of binary NN
collisions {\em without} modification. Thus a $R_{AA}$
 smaller(larger) than unity means
suppression(enhancement) due to medium effect. At RHIC this ratio
at large $p_t>6GeV$ has been measured to be a constant, about
$0.2$ for the most central AuAu collisions. Accurate calibration
of hard processes in pp and dAu collisions, as well as with hard
photon measurements (which show no quenching)
\cite{rhic_white_paper} resulted in quite accurate knowledge of
jet production geometry, for any impact parameter $b$ (or
centrality bins, often characterized by the number of nucleon
participants $N_{part}$ in a collision event).  While quenching is
firmly established as a final state effect,  many efforts to
understand its microscopic mechanism are not yet conclusive. Those
include pQCD gluon radiation with Landau-Pomeranchuk-Migdal (LPM)
effect \cite{Baier:1996sk}, synchrotron-like radiation on coherent
fields \cite{Shuryak:2002ai,Kharzeev:2008qr}, elastic scattering
loss
 \cite{Wicks:2005gt}, etc.
The fate of deposited energy was discussed in
Refs\cite{Stoecker:2004qu,CasalderreySolana:2004qm}
led to predictions of
``conical flow'' correlated with
experimentally observed conical structures in correlations involving 2 or 3
particles,  for reviews see e.g.
\cite{Gyulassy:2003mc,Shuryak_08}.

{\em Jet tomography and the geometric limit.---}  In non-central
collisions the overlap region
 of two colliding nuclei
 has almond-like shape: thus jets penetrating
 the fireball in different directions lose different
amount of energy according to their varying paths. Their
yield distribution $d^2N/dp_td\phi$ in azimuthal angle $\phi$
 (with respect to the reaction plane) for  high $p_t$ hadrons
  thus provides a ``tomography'' of the
fireball\cite{Gyulassy:2000gk,Wang:2000fq,Shuryak:2001me}. We will
focus on the second Fourier coefficient
\begin{equation}
 v_2(p_t,b)\equiv \frac{\int_0^{2\pi}d\phi\, \cos(2\phi)\,[d^2N/dp_td\phi]}
 {\int_0^{2\pi}d\phi \, [d^2N/dp_td\phi]}
\end{equation}
 depending on  impact parameter $b$ for large $p_t> 6\, {\rm GeV}$ where
 hard processes dominate and  dependence on
$p_t$ is weak\cite{phenix_raa_v2}.

 Unexpectedly, measured $ v_2(p_t,b)$ happen to be
considerably larger than what jet quenching models predicted. The
aim of our work is to provide simultaneous description of both
$R_{AA}$ and $v_2$ at high $p_t$ based on theoretically known
geometry of jet production and bulk matter evolution. One
important concept of the analysis is the so called {\em geometric
limit}, first suggested by one of us in \cite{Shuryak:2001me}: the
observed asymmetry should be less than some value
  $ v_2(large \, p_t,b) < v_2^{max}(b) $ provided by the geometry
of the overlap region of two colliding nuclei. The  idea
\cite{Shuryak:2001me} was that for very strong quenching
 only jets emitted from the surface of the almond can be observed.
 Two other  assumptions were made, namely:
(i) quenching is proportional to  matter density;
(ii) colliding nuclei were approximated  by homogeneous
sharp-edge spheres. However even early
 experimental data showed that $v_2$ is actually well $above$
this bound. Subsequent studies by Drees,{\it et al}
\cite{Drees:2003zh} relaxed the second assumption, with realistic
nuclear shapes, which only made contradiction with data even
stronger (see e.g. their Fig.3(d)).

The main lesson from those studies is that quenching is $not$
proportional to the matter density, but a nontrivial function of
it. Assuming some  form of this function,
 one can then calculate both observables
$v_2(b)$ and $R_{AA}$.

{\em Layer-wise geometrical limit.---} Systematically slicing the
(expanding) fireball into shells with the entropy density
 $s_a<s\le s_b$, we calculate what $R_{AA}(b)$ and
$v_2(b)$ would result with such single shell being the sole source
of quenching by a Glauber simulation of AuAu collisions and jet
production as in \cite{Shuryak:2001me,Drees:2003zh}. With the
quenching function $\kappa(s)$ assumed to be concentrated at this
slice $\kappa_{ab}\cdot \theta(s-s_a)\cdot\theta(s_b-s)$, the
distribution in survival probability $f$ can be calculated and
directly leads to evaluation of $R_{AA}$:
\begin{equation}\label{eqn_rf}
f = e^{-\int_{path} \kappa[s(l)] \, s(l) \, l\, dl}\,\, , \,
R_{AA}=<f^{n-2}> \,\, , \, n\approx 8.10
\end{equation}
Extra $l$ in the path corresponds to radiative LPM theory
\cite{Baier:1996sk}. The power index $n$ comes from the $\pi_0$
$p_t$ spectrum in pp collisions, see detailed discussions in
\cite{phenix_raa_v2}. For each density shell the absorption
coefficient $\kappa_{ab}$ (in unit $fm$) is then fixed by $R_{AA}$
data \cite{phenix_raa_v2} parameterized by
$R_{AA}(p_T>5GeV)=[1-8.3\cdot 10^{-3} \cdot
N_{part}^{0.58}]^{n-2}$. Then we calculate $v_2$, by sampling
half of the jets travelling in $x$ directions $\pm 5^o$ and the
other half in $y$ direction and extracting the difference in the
respective $R_{AA}^{x(y)}$\cite{phenix_raa_v2}. For the Glauber
initial condition we follow hydro calculations (see e.g.
\cite{Teaney:2001av}) to scale entropy density with local
participant density, and for bulk evolution we use 1-D Bjorken
dilution which is appropriate till time $\sim 10\,fm/c$ (see e.g.
\cite{Kolb:2003gq}). Jet production points are simulated according
to binary collision density. We have  24 entropy shells,
(0,1],(1,2],...,(23,24] (in $/fm^3$ units).
\begin{figure*}
  \hskip 0in\includegraphics[width=8.cm]{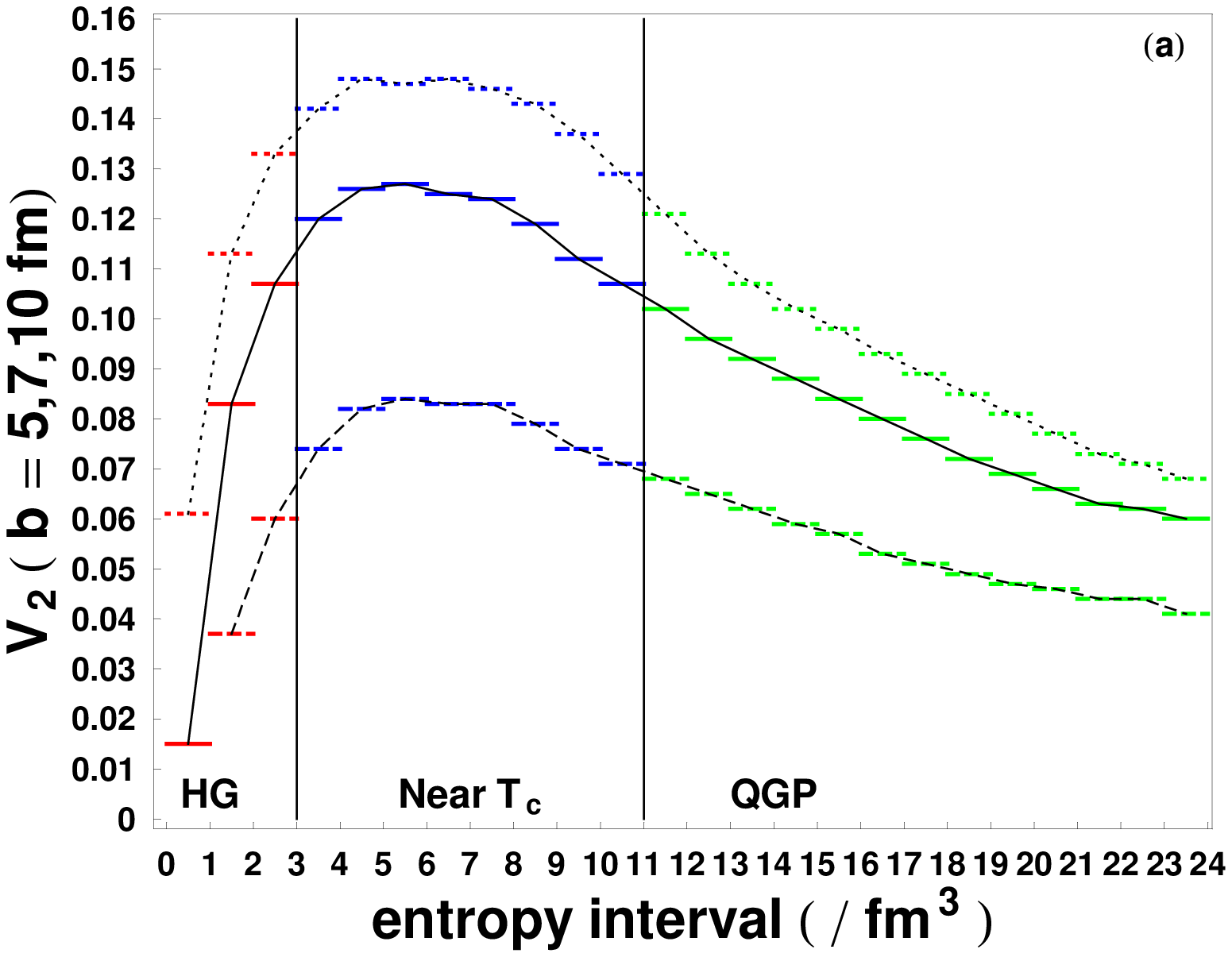}
    \hskip 0.1in\includegraphics[width=7.2cm]{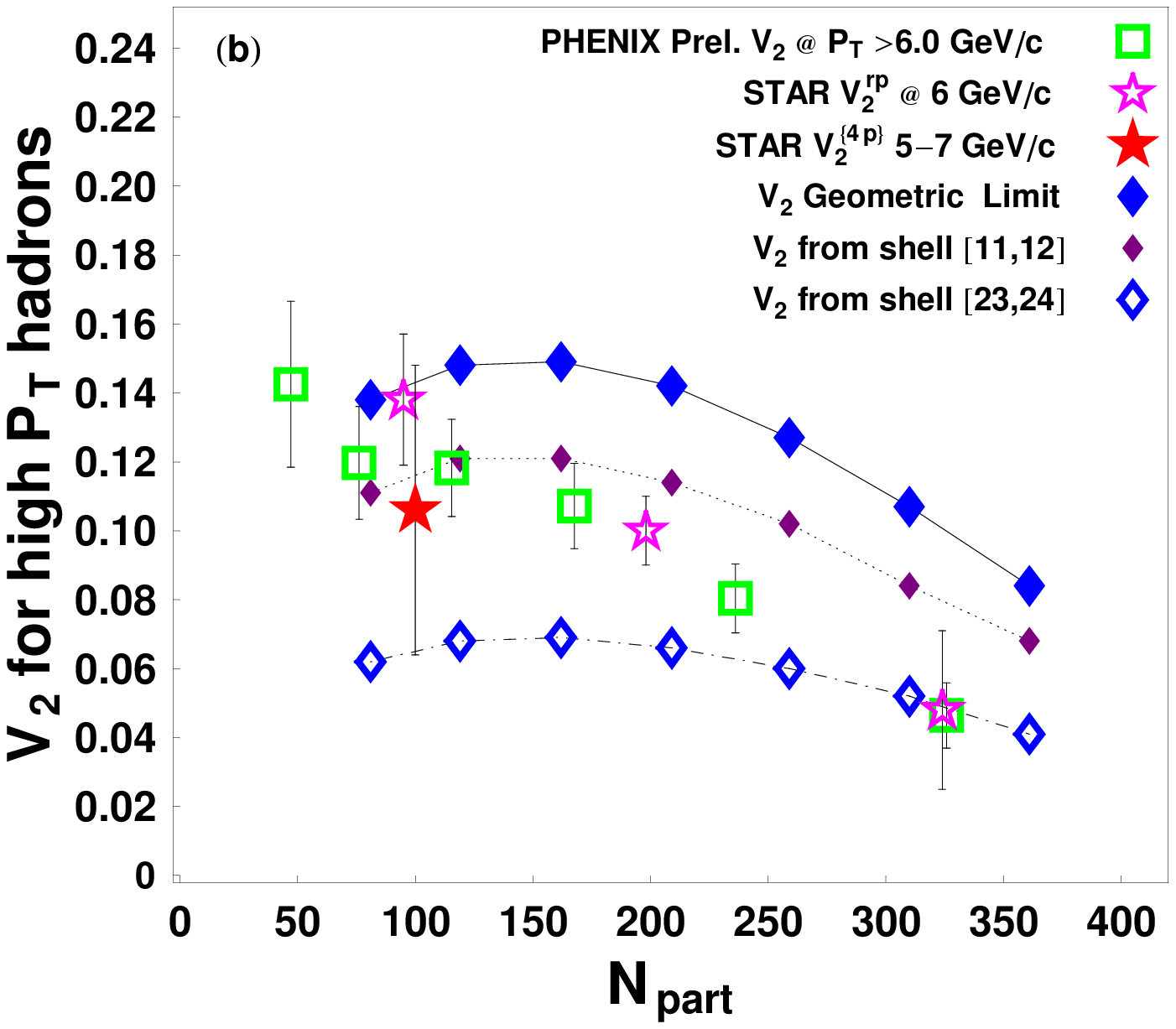}
 \caption{\label{fig_v2}(color online) (\textbf{a}) The $v_2$ obtained for each entropy shell
 at $b=5fm$(dashed),$7fm$(solid), and $10fm$(dotted) respectively; (\textbf{b}) $v^{max}_2$ for high $p_T$
 hadrons calculated at different $N_{part}$ as compared with
 available RHIC data from \protect\cite{phenix_raa_v2,Rui_Wei} and \protect\cite{star_v2}.  }
 \end{figure*}

The resulting  $v_2$ for three impact parameters $b=5,7,10 \, fm $
(bottom-to-top) are shown in Fig.\ref{fig_v2}(\textbf{a}). \\
(i) Note  that certain entropy shells produce $v_2$
much $larger$ than the old geometric limits of
refs.\cite{Shuryak:2001me,Drees:2003zh},
  corresponding to surface emission (small $s$ at the left side of the plot).\\
(ii) the existence of the
$maximum$ $v^{max}_2(b)$ leads to {\it layer-wise
geometrical limit}: its dependence on centrality is shown in
Fig.\ref{fig_v2}(\textbf{b}) by filled big blue diamonds.\\
(iii) Interestingly enough, the entropy shells where the maxima
occur (for all centralities)  correspond to the same interval
$s=4-8 \,fm^{-3}$, which  is in fact quite special: it corresponds
$exactly$ to the vicinity of the QCD phase transition (see e.g.
\cite{Cheng:2007jq}). These curves reflect not only the  geometry
of the respective entropy shells, but also their placement
relative to the jet production points.

After these studies of single shells, we turn to the compiled
high-$p_t$ RHIC data on $v_2(b)$, shown in
Fig.\ref{fig_v2}(\textbf{b}). We include only data for ``hard''
hadrons with $p_t>6\,GeV$ from PHENIX (open green boxes) and STAR
(open magenta stars) collaborations. Comparing these data points
to our {\em layer-wise geometric limit} (filled big blue
diamonds), we do observe  that  all the data points are (within
error bars) indeed
 $below$ this proposed bound. We also show $v_2(b)$ lines
which would come out if all jet quenching would be due to two
other single entropy shells, with $s= (11,12] fm^{-3}$ (filled
small purple diamonds) and  $s=(23,24] fm^{-3}$ (open blue
diamonds). Those correspond to the QGP phase, near and far from
the transition region: the values of  $v_2(b)$ from those shells
are significantly smaller than the maximal. Now we qualitatively
understand the experimental trend:  going from the more central to
the more peripheral collisions,  quenching geometry shifts from
quenching at high  density shells (QGP), to the near-$T_c$ region
at $N_p\sim 100$ (approaching the upper limit). For extremely
peripheral collisions we expect $v_2$ to decrease again,
reflecting geometry of the low entropy density shells (the
hadronic phase).

\begin{figure}
  \hskip 0in\includegraphics[width=8.cm]{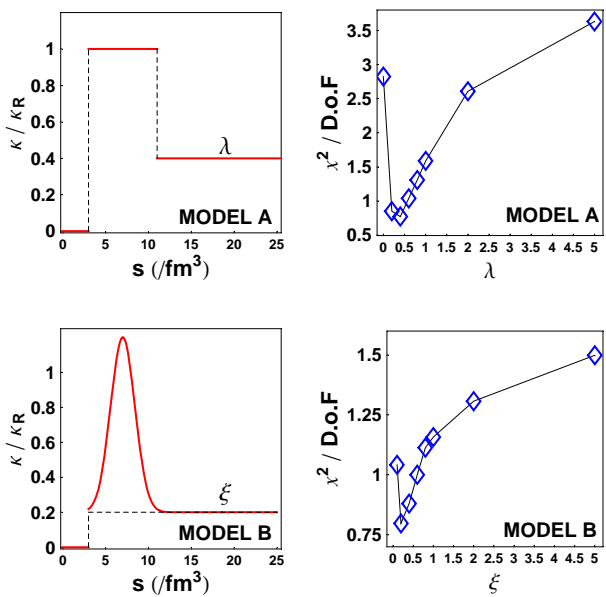}
  \vskip 0.1in
 \caption{\label{fig_model_chis}(color online) (left) Schematic demonstration of the
 quenching functions of our Model A and B; (right) The $\chi^2/D.o.F$ when fitting
 the $v_2$ data with different values of parameters $\lambda$ ($\xi$) in our Model A (B), see text.   }
 \end{figure}

{\em Modelling tomography  of jet quenching.--- } 
 We now turn from individual shells
to  realistic models, describing the combined effect of all of them.\\
{\bf Model A ---} a two-phase scenario model, in which we assume
the quenching function $\kappa(s)$ with two parameters: one
 in the near-$T_c$ region and the other for the
QGP phase, i.e.
\begin{equation}\label{two_phase}
\kappa(s)=\kappa_R \times [1\cdot \theta(s-s^c_1)\cdot
\theta(s^c_2-s) + \lambda\cdot \theta(s-s^c_2)]
\end{equation}
with $s^c_1=3/fm^3$ and $s^c_2=11/fm^3$ bracketing the near-$T_c$
region. The parameter $\kappa_R$ is  globally fitted from
$R_{AA}(N_{part})$ (for each given $\lambda$), while $\lambda$
characterizes the relative quenching strength between the
near-$T_c$ region and the QGP, with its best value to be
determined from a global fitting for $v_2(N_{part})$.\\
{\bf Model B ---} a scenario featuring peaked quenching strength
at $T_c$, which assumes
\begin{equation}\label{bump}
\kappa(s)=\kappa_R \times [e^{-(\frac{s-s_c}{s^c_w})^2}\, \cdot
\theta(s-s^c_1) + \xi \cdot \theta(s-s^c_1)]
\end{equation}
with $s^c=7/fm^3$ and $s^c_w=2/fm^3$ spanning the near-$T_c$
region according to lattice results\cite{Cheng:2007jq}.

Schematic sketches of the two models' $\kappa$ are shown in
Fig.\ref{fig_model_chis}(left) and $\chi^2/D.o.F$ from fitting the
$v_2$ data (both the PHENIX and the STAR points), with a variety
of choices of $\lambda$ (Model A) / $\xi$ (Model B), are shown in
Fig.\ref{fig_model_chis}(right). The plots suggest that current
$v_2$ data favors the relative quenching strength $\lambda=0.4$
for Model A and $\xi=0.2$ for Model B, {\em both favoring a
scenario that in relativistic heavy ion collisions the jets are
quenched about 2-5 times stronger in the near-$T_c$ region than
the higher-T QGP phase.}

\begin{figure}
    \hskip 0.1in\includegraphics[width=7.7cm]{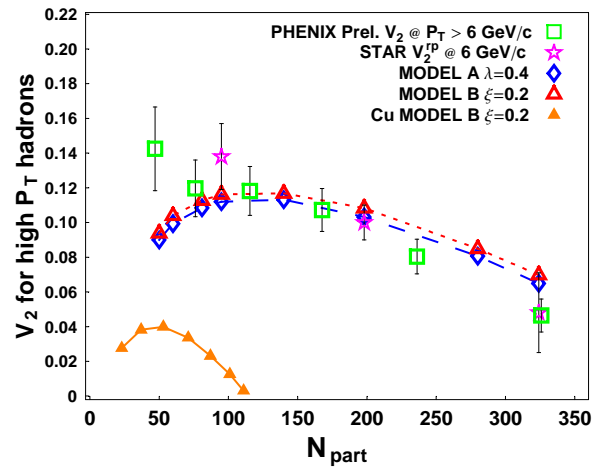}
 \caption{\label{fig_model_v2}(color online) Comparison between
 $v_2$ experimental data and $v_2$ calculated from our models, see text.}
 \end{figure}

We also plot in Fig.\ref{fig_model_v2} the $v_2(N_{part})$
obtained with the above optimal parameters: Model A with
$\kappa_R=0.00435 fm$ and $\lambda=0.4$, Model B with
$\kappa_R=0.00745fm$ and $\xi=0.2$. Both of them describe current
data very well and predict rapid dropping of $v_2$ at the very
peripheral end $N_p\ll 100$.

{\em Conclusions and discussion.--- } We started  with
the calculation of the
 ``layer-wise geometric limit'' for
 models describing jet quenching \be
v_2(b)< v_2^{max}(b)\ee
 where the r.h.s. is shown by the filled big blue diamonds in Fig.\ref{fig_v2}(\textbf{b})
 and corresponds to particular density shells.
Unlike previously proposed bounds, this one is indeed satisfied by all available data (for
large enough $p_t$ ,  within error bars). The limit
can be reached only when the jet quenching is overwhelmingly
dominated by the matter shells with the entropy density  $s=4-8 \,fm^{-3}$ since
only those have the right geometrical properties: the data points
suggest this seems indeed to be the case for the AuAu collisions
at RHIC at $N_p\sim 100$.

While previous models\cite{Shuryak:2001me,Drees:2003zh}
 failed to reproduce the high $p_t$ $v_2$ and $R_{AA}$
simultaneously, we now presented two models which can do so. The
key is the nontrivial dependence of quenching on the (entropy)
density. We concluded that the angular dependence of jet quenching
 indicates its strong enhancement near the QCD phase transition,
 about several times stronger than in the QGP.

Why can it be so? Perhaps a near-$T_c$ peak in jet quenching
should not be too surprising, as we already saw similar
peaks/sharp-valleys around $T_c$ for other properties of QGP, from
trace anomaly, specific heat and speed of sound\cite{Cheng:2007jq}
to shear and bulk viscosities\cite{Csernai:2006zz}. Recently the
jet quenching strength was found to inversely related to shear
viscosity in weakly coupled QGP\cite{Majumder:2007zh} --- such
relation if naively extrapolated and combined with the minimum of
shear viscosity at $T_c$ would also point to a near-$T_c$ peak of
jet quenching. It was also proposed in \cite{Pantuev:2005jt} that
switching on quenching only after a {\em global} time $\tau_q\sim
2\, fm$ one can obtain better values of the asymmetry: such effect
is incorporated by near-$Tc$ dominance in a much more plausible
manner via {\em local} density evolution.

A microscopic explanation may be provided by recent magnetic
scenario for the near-$T_c$ QCD plasma, in which this narrow
T-region is treated as a {\bf magnetic plasma} of light
monopoles\cite{Liao:2006ry}. In the same region quarks/gluons are
few times heavier and thus get less energy for the same momentum
transfer. When a fast electric charge (the jet) penetrates such
plasma, its strong transverse magnetic field easily accelerates
the abundant light monopoles into an overheated magnetic ``coil''
behind it via the dual Faraday effect, leading to substantial
energy loss of the jet \cite{Liao:2005pa}.

It will be interesting to extend the present study to different
colliding nuclei A and beam energy $\sqrt{s}$: the data are
becoming available (see e.g.\cite{Phenix_more}) and the phenomena
are rich as the jet production, the bulk evolution, and the pp
reference all scale differently with A and $\sqrt{s}$. In
Fig.\ref{fig_model_v2} we have included the prediction for high
$p_t$ $v_2$ of CuCu 200GeV collisions from our model B fixed by
AuAu (orange filled triangle), to be tested by data. More
dedicated studies (including different initial scaling, different
path length dependence, etc) will be reported in \cite{LS_coming}.

{\textbf{Note added:}}  After the Letter was submitted, PHENIX
run7 preliminary data were released\cite{Rui_Wei}. They are now
included in Fig.\ref{fig_v2}(a) and Fig.\ref{fig_model_v2}
(squares): as one can see they agree with our model well. As also
shown in \cite{Rui_Wei}, most other models of quenching give $v_2$
2-3 times smaller than data.

This work was supported in parts by the US-DOE grant
DE-FG-88ER40388. JL is also supported by the Director, Office of
Energy Research, Office of High Energy and Nuclear Physics,
Divisions of Nuclear Physics, of the U.S. Department of Energy
under Contract No. DE-AC02-05CH11231. The authors are grateful to
Barbara Jacak for valuable discussions and to Rui Wei for help on
data. JL also thanks V. Koch, P. Jacobs, J. Jia, R. Lacey, and L.
McLerran for helpful discussions.

\bibliographystyle{apsrev}

\begin{thebibliography}{99}

\bibitem{Shuryak:1978ij}
  E.~V.~Shuryak,
  Phys.\ Lett.\  B {\bf 78}, 150 (1978)
  [Sov.\ J.\ Nucl.\ Phys.\  {\bf 28}, 408.1978\ YAFIA,28,796 (1978)].

\bibitem{Liao:2006ry}
  J.~Liao and E.~Shuryak,
  Phys.\ Rev.\  C {\bf 75}, 054907 (2007);
    Phys.\ Rev.\ Lett.\  {\bf 101}, 162302 (2008).
  M.~N.~Chernodub and V.~I.~Zakharov,
  Phys.\ Rev.\ Lett.\  {\bf 98}, 082002 (2007).

\bibitem{Bjorken:1982tu}
  J.~D.~Bjorken,
  FERMILAB-PUB-82-059-THY.



\bibitem{Appel_Blaizot_McLerran}
D.~A.~Appel,
Phys.\ Rev.\ D {\bf 33}, 717 (1986);
J.~P.~Blaizot and L.~D.~McLerran,
Phys.\ Rev.\ D {\bf 34}, 2739 (1986).

\bibitem{Gyulassy:1990ye}
  M.~Gyulassy and M.~Plumer,
  Phys.\ Lett.\  B {\bf 243}, 432 (1990);
  X.~N.~Wang and M.~Gyulassy,
  Phys.\ Rev.\ Lett.\  {\bf 68}, 1480 (1992).

\bibitem{rhic_white_paper}
  J.~Adams {\it et al.},
  Nucl.\ Phys.\  A {\bf 757}, 102 (2005).
  K.~Adcox {\it et al.},
  Nucl.\ Phys.\  A {\bf 757}, 184 (2005).




\bibitem{Baier:1996sk}
  R.~Baier, Y.~L.~Dokshitzer, A.~H.~Mueller, S.~Peigne and D.~Schiff,
  Nucl.\ Phys.\  B {\bf 484}, 265 (1997).

\bibitem{Shuryak:2002ai}
  E.~V.~Shuryak and I.~Zahed,
  Phys.\ Rev.\  D {\bf 67}, 054025 (2003)
  [arXiv:hep-ph/0207163].

\bibitem{Kharzeev:2008qr}
  D.~E.~Kharzeev,
  arXiv:0806.0358 [hep-ph].

\bibitem{Wicks:2005gt}
  S.~Wicks, W.~Horowitz, M.~Djordjevic and M.~Gyulassy,
  Nucl.\ Phys.\  A {\bf 784}, 426 (2007).


\bibitem{Stoecker:2004qu}
  H.~Stoecker,
  Nucl.\ Phys.\  A {\bf 750}, 121 (2005).

\bibitem{CasalderreySolana:2004qm}
  J.~Casalderrey-Solana, E.~V.~Shuryak and D.~Teaney,
  J.\ Phys.\ Conf.\ Ser.\  {\bf 27}, 22 (2005)
  [arXiv:hep-ph/0411315].




\bibitem{Gyulassy:2003mc}
M.~Gyulassy, I.~Vitev, X.~N.~Wang and B.~W.~Zhang,
  arXiv:nucl-th/0302077;
  X.~N.~Wang,
  Nucl.\ Phys.\  A {\bf 750}, 98 (2005);
  J.~Casalderrey-Solana and C.~A.~Salgado,
  Acta Phys.\ Polon.\  B {\bf 38}, 3731 (2007).

\bibitem{Shuryak_08}  E.~Shuryak,
  Prog.\ Part.\ Nucl.\ Phys.\  {\bf 62}, 48 (2009).



\bibitem{Gyulassy:2000gk}
  M.~Gyulassy, I.~Vitev and X.~N.~Wang,
  Phys.\ Rev.\ Lett.\  {\bf 86}, 2537 (2001).

  \bibitem{Wang:2000fq}
  X.~N.~Wang,
  Phys.\ Rev.\  C {\bf 63}, 054902 (2001).


\bibitem{Shuryak:2001me}
  E.~V.~Shuryak,
  Phys.\ Rev.\  C {\bf 66}, 027902 (2002).

\bibitem{phenix_raa_v2}
  S.~S.~Adler {\it et al.},  [PHENIX Collaboration],
  Phys.\ Rev.\  C {\bf 76}, 034904 (2007);
 Phys.\ Rev.\ Lett.\  {\bf 101}, 232301 (2008);
  arXiv:0903.4886 [nucl-ex].



\bibitem{Drees:2003zh}
  A.~Drees, H.~Feng and J.~Jia,
  Phys.\ Rev.\  C {\bf 71}, 034909 (2005)
  [arXiv:nucl-th/0310044].






\bibitem{Teaney:2001av}
  D.~Teaney, J.~Lauret and E.~V.~Shuryak,
  Phys.\ Rev.\ Lett.\  {\bf 86}, 4783 (2001);
  arXiv:nucl-th/0110037.


\bibitem{Kolb:2003gq}
  P.~F.~Kolb,
  Heavy Ion Phys.\  {\bf 21}, 243 (2004).





\bibitem{Rui_Wei}
R. Wei, to appear in Proceedings of Quark Matter 2009.

\bibitem{star_v2}
J.~Adams {\it et al.}  [STAR Collaboration],
  Phys.\ Rev.\ Lett.\  {\bf 93}, 252301 (2004);
  K.~Filimonov
  Nucl.\ Phys.\  A {\bf 715}, 737 (2003);
  R.~Snellings,
  Heavy Ion Phys.\  {\bf 21}, 237 (2004).



\bibitem{Cheng:2007jq}
  M.~Cheng {\it et al.},
  Phys.\ Rev.\  D {\bf 77}, 014511 (2008).



\bibitem{Csernai:2006zz}
  L.~P.~Csernai, J.~I.~Kapusta and L.~D.~McLerran,
  Phys.\ Rev.\ Lett.\  {\bf 97}, 152303 (2006).
  F.~Karsch, D.~Kharzeev and K.~Tuchin,
  Phys.\ Lett.\  B {\bf 663}, 217 (2008).



\bibitem{Majumder:2007zh}
  A.~Majumder, B.~Muller and X.~N.~Wang,
  Phys.\ Rev.\ Lett.\  {\bf 99}, 192301 (2007).

\bibitem{Pantuev:2005jt}
  V.~S.~Pantuev,
  JETP Lett.\  {\bf 85}, 104 (2007).

\bibitem{Liao:2005pa}
  J.~Liao and E.~V.~Shuryak,
  Phys.\ Rev.\  D {\bf 73}, 014509 (2006);
  Nucl.\ Phys.\  A {\bf 775}, 224 (2006);
  Phys.\ Rev.\  C {\bf 77}, 064905 (2008);
  arXiv:0804.4890 [hep-ph].

\bibitem{Phenix_more}
A.~Adare {\it et al.} [PHENIX Collaboration],
  Phys.\ Rev.\ Lett.\ {\bf 101}, 162301 (2008);
  Phys.\ Rev.\ Lett.\ {\bf 98}, 162301 (2007);
  Phys.\ Rev.\ Lett.\ {\bf 94}, 232302 (2005).

\bibitem{LS_coming}
 J.~Liao and E.~Shuryak, in preparation.

\end{thebibliography}

\end{document}